\documentclass{ecai} 

\usepackage{latexsym}
\usepackage{amssymb}
\usepackage{amsmath}
\usepackage{amsthm}
\usepackage{booktabs}
\usepackage{enumitem}
\usepackage{graphicx}
\usepackage{color}
\usepackage{booktabs}
\usepackage{multirow}
\usepackage{graphicx}
\usepackage{adjustbox}
\usepackage{array}

\newcommand{\BibTeX}{B\kern-.05em{\sc i\kern-.025em b}\kern-.08em\TeX}

\begin{document}


\begin{frontmatter}


\paperid{123} 


\title{MixGAN: A Hybrid Semi-Supervised and Generative Approach for DDoS Detection in Cloud-Integrated IoT Networks}


\author[A]{\fnms{Tongxi}~\snm{Wu}\footnote{Equal contribution.}}
\author[A]{\fnms{Chenwei}~\snm{Xu}\footnotemark}
\author[A,B]{\fnms{Jin}~\snm{Yang}\thanks{Corresponding Author Jin Yang. Email: yangjin66@scu.edu.cn.}\footnote{This paper has been accepted at the Main Conference of ECAI 2025.}}

\address[A]{College of Cyber Science and Engineering, Sichuan University, Chengdu 610207, China}
\address[B]{College of Information Science and Technology, Tibet University, Lhasa 850000, China}


\begin{abstract}
The proliferation of cloud-integrated IoT systems has intensified exposure to Distributed Denial of Service (DDoS) attacks due to the expanded attack surface, heterogeneous device behaviors, and limited edge protection. However, DDoS detection in this context remains challenging because of complex traffic dynamics, severe class imbalance, and scarce labeled data. While recent methods have explored solutions to address class imbalance, many still struggle to generalize under limited supervision and dynamic traffic conditions. To overcome these challenges, we propose MixGAN, a hybrid detection method that integrates conditional generation, semi-supervised learning, and robust feature extraction. Specifically, to handle complex temporal traffic patterns, we design a 1-D WideResNet backbone composed of temporal convolutional layers with residual connections, which effectively capture local burst patterns in traffic sequences. To alleviate class imbalance and label scarcity, we use a pretrained CTGAN to generate synthetic minority-class (DDoS attack) samples that complement unlabeled data. Furthermore, to mitigate the effect of noisy pseudo-labels, we introduce a MixUp-Average-Sharpen (MAS) strategy that constructs smoothed and sharpened targets by averaging predictions over augmented views and reweighting them towards high-confidence classes. Experiments on NSL-KDD, BoT-IoT, and CICIoT2023 demonstrate that MixGAN achieves up to 2.5\% higher accuracy and 4\% improvement in both TPR and TNR compared to state-of-the-art methods, confirming its robustness in large-scale IoT-cloud environments. The source code is publicly available at \url{https://github.com/0xCavaliers/MixGAN}.
\end{abstract}

\end{frontmatter}


\section{Introduction}
Cloud computing provides on-demand access to resources such as servers, storage, networks~\cite{chen2024multicenter}, and AI services~\cite{zeng2024implementation}, improving data processing efficiency via virtualization. Services like SaaS, PaaS, and IaaS~\cite{mohan2024securing} offer scalable, flexible, and cost-effective solutions. However, this increased reliance also introduces critical data and privacy risks~\cite{chang2015towards, hizal2024novel, kim2024zero}.

\begin{figure}[!ht]
\centering
\includegraphics[width=0.45\textwidth]{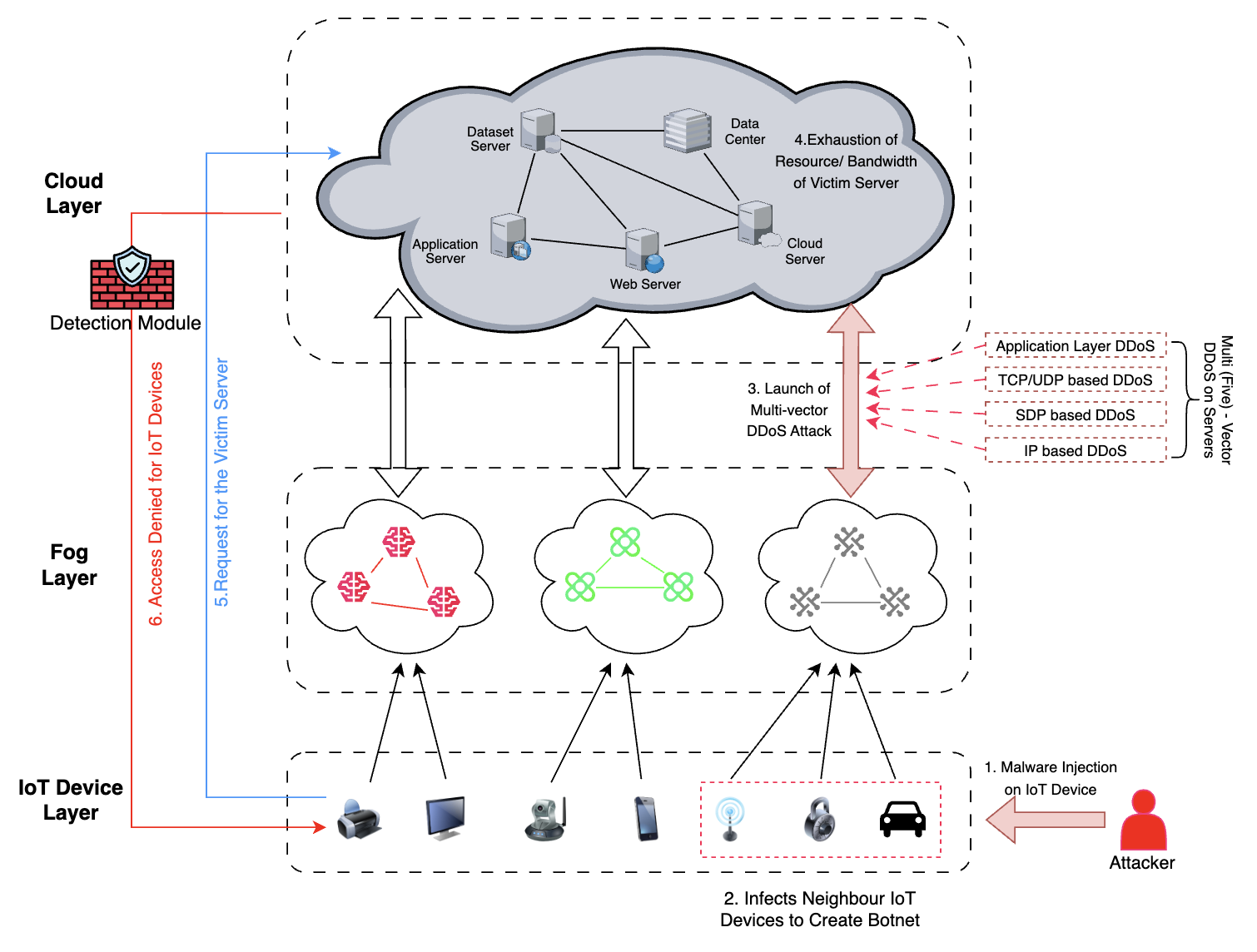}
\caption{
\scriptsize Overview of a DDoS attack scenario in a cloud-integrated IoT environment.
}
\label{fig:simulation_architecture}
\end{figure}

Among emerging threats, Distributed Denial of Service (DDoS) attacks stand out for their scale and impact~\cite{cai2023adam, kumari2023comprehensive}. Coordinated attacks launched from thousands of compromised devices can exhaust system resources, disrupt services, and are increasingly amplified through cloud infrastructures~\cite{hnamte2024ddos, agrawal2019defense}. Malware-as-a-Service (MaaS) and Attack-as-a-Service (AaaS) models further lower the barrier to executing such attacks. These attacks are often launched from compromised IoT devices and propagate across layered infrastructures toward centralized cloud services. Figure~\ref{fig:simulation_architecture} provides an architectural overview of such attack scenarios in a cloud-integrated IoT environment, including traffic sources, target servers, and the placement of the detection module.

The proliferation of cloud-integrated IoT systems has introduced new vulnerabilities to Distributed Denial of Service (DDoS) attacks~\cite{hizal2024novel, hnamte2024ddos}, driven by heterogeneous device behaviors~\cite{kumar2024comprehensive}, large-scale connectivity~\cite{aguru2024lightweight}, and weak edge defenses~\cite{mangala2024secure}. These attacks disrupt system availability by overwhelming cloud resources with massive illegitimate traffic, posing significant threats to smart cities, healthcare, and industrial infrastructures.

Traditional signature-based defenses struggle against these evolving threats~\cite{manzil2024detection, birthriya2025detection}. Consequently, AI and deep learning techniques have been widely adopted to enhance DDoS detection~\cite{kumar2024comprehensive, aguru2024lightweight}. Models such as KNN, SVM, Decision Trees, CNNs, LSTMs, and RNNs~\cite{bala2024ai, hekmati2024correlation} have achieved success in identifying attacks and anomalies. Recent work also explores hybrid models and semi-supervised learning~\cite{mvula2024survey} to tackle label scarcity and resource constraints in IoT-cloud environments.

Despite these advances, DDoS detection in dynamic, large-scale IoT-cloud systems remains highly challenging, especially under severe class imbalance and limited labeled data~\cite{naeem2023federated}. To address these challenges, this paper proposes a novel DDoS detection framework based on generative augmentation and semi-supervised optimization. The main contributions are summarized as follows:

\begin{itemize}
    \item We design an improved 1-D WideResNet~\cite{he2016deep, zagoruyko2016wide} with expanded network width to efficiently capture complex temporal traffic patterns. Using 1-D convolutional kernels, the model is tailored to sequence-based traffic data, achieving a better trade-off between expressive capacity and computational efficiency than traditional CNNs or LSTMs, making it suitable for real-time DDoS detection in dynamic IoT-cloud environments.
    
    \item We integrate a CTGAN-based conditional sample generator~\cite{xu2019modeling, habibi2023imbalanced} into the semi-supervised learning framework. Beyond simple data augmentation, CTGAN generates diverse, class-conditional synthetic traffic variants that complement unlabeled samples, enabling more accurate pseudo-labeling and improved label propagation under severe class imbalance and label scarcity.
    
    \item We propose MAS (MixUp-Average-Sharpen), a MixMatch-inspired~\cite{berthelot2019mixmatch} semi-supervised strategy tailored for traffic anomaly detection. MAS generates multiple CTGAN-conditioned variants for each unlabeled sample, predicts their pseudo-labels, then averages and sharpens them into confident soft targets. These are incorporated into a MixUp-based training scheme with labeled data, enhancing stability, generalization, and robustness against diverse and evolving DDoS patterns.
\end{itemize}


\section{Related Work}

\subsection{Intrusion Detection Systems in Cloud and IoT Environments}

Deep learning models have been widely adopted to enhance intrusion detection systems (IDS) in IoT-cloud environments. Doriguzzi-Corin et al.~\cite{doriguzzi2020lucid} proposed LUCID, a lightweight CNN-based framework for efficient and explainable DDoS detection under resource-constrained settings. Velliangiri et al.~\cite{velliangiri2020fuzzy} proposed FT-DBN, which optimized deep belief networks using fuzzy-Taylor series expansion and elephant herd optimization for DDoS detection. Emil Selvan et al.~\cite{gsr2023facvo} introduced FACVO-DNFN, combining fractional anti-corona virus optimization with a deep neuro-fuzzy network to improve feature fusion and classification. Dehghani et al.~\cite{dehghani2022hybrid} developed HLBO+DSA by integrating hybrid leader-based optimization with deep stacked autoencoders, while Balasubramaniam et al.~\cite{balasubramaniam2023optimization} further extended this approach with GHLBO+DSA to enhance convergence and detection robustness. 

Recent works have explored graph-based and federated architectures, such as the self-supervised temporal contrastive graph model TCG-IDS~\cite{wu2025tcg}, the semi-supervised federated Cycle-Fed framework~\cite{xiao2024cycle}, and the hierarchical ensemble GNN FTG-Net-E~\cite{bakar2024ftg}. While promising, these methods incur substantial computational and memory costs, limiting their practicality for real-time IoT detection. Other approaches still rely heavily on supervised learning and oversampling~\cite{habibi2023imbalanced}, which struggle to generalize under dynamic, label-scarce conditions. Recent surveys~\cite{bala2024ai, mvula2024survey} emphasize the potential of semi-supervised learning and generative augmentation in cybersecurity, yet adoption in IoT-cloud IDS remains limited. Moreover, DiffuPac~\cite{bin2024diffupac} shows that packet-level adversarial traffic generated by diffusion models can evade state-of-the-art NIDS by an average of 6.69 percentage points in black-box settings, underscoring the need for robust semi-supervised detectors resilient to stealthy perturbations.

\subsection{GAN-based Methods for Traffic Classification and Intrusion Detection}

Recent studies have applied generative adversarial networks (GANs) to improve intrusion detection, particularly under class imbalance and limited data. For example, GAN-AE~\cite{boppana2023gan} combines adversarial training with autoencoder-based feature extraction for MQTT networks, while CL-GAN~\cite{li2024hda} integrates CNN-LSTM architectures to enhance anomaly detection. AE-WGAN~\cite{seghair2024continual} leverages denoising autoencoders with Wasserstein GANs for better feature representation, and SYN-GAN~\cite{rahman2024syn} generates synthetic traffic to address data scarcity. S2CGAN-IDS~\cite{wang2023effective} augments minority classes in both data and feature spaces for IoT security, whereas TMG-GAN~\cite{ding2023tmg} employs multiple generators to improve diversity and quality of synthetic attack samples.

While these GAN-based approaches improve minority class detection and anomaly identification, they often face challenges related to mode collapse, training stability, or increased model complexity, limiting their scalability in large-scale IoT-cloud environments. 

\section{Lightweight NS3-Based Cloud-IoT Simulation Framework}

\begin{figure}[!ht]
\centering
\includegraphics[width=2.0in]{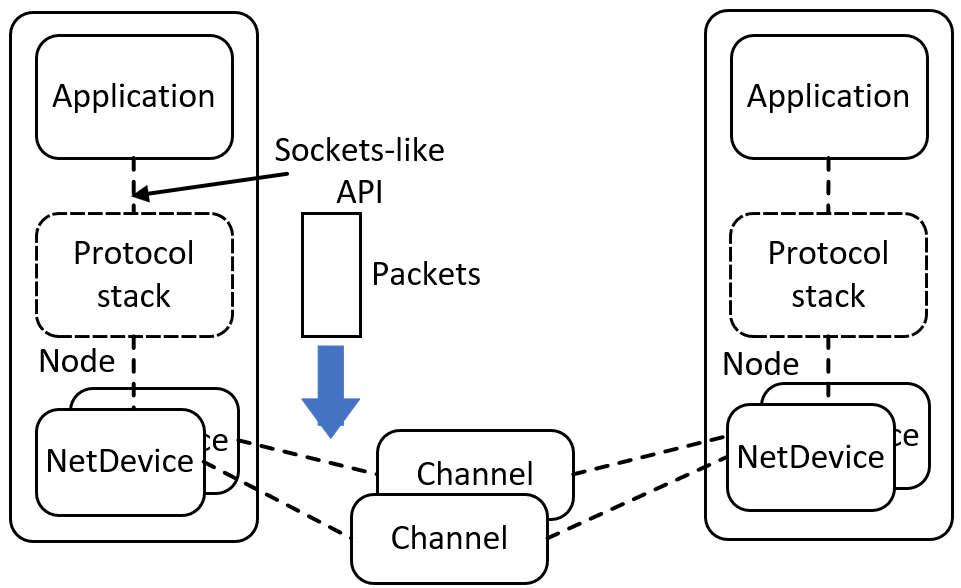}
\caption{
\scriptsize NS3 Simulation Network Architecture: Illustrating the node components, protocol stack, and point-to-point communication channels.
}
\label{fig:architecture}
\end{figure}

\begin{figure}[!ht]
\centering
\includegraphics[width=3.2in]{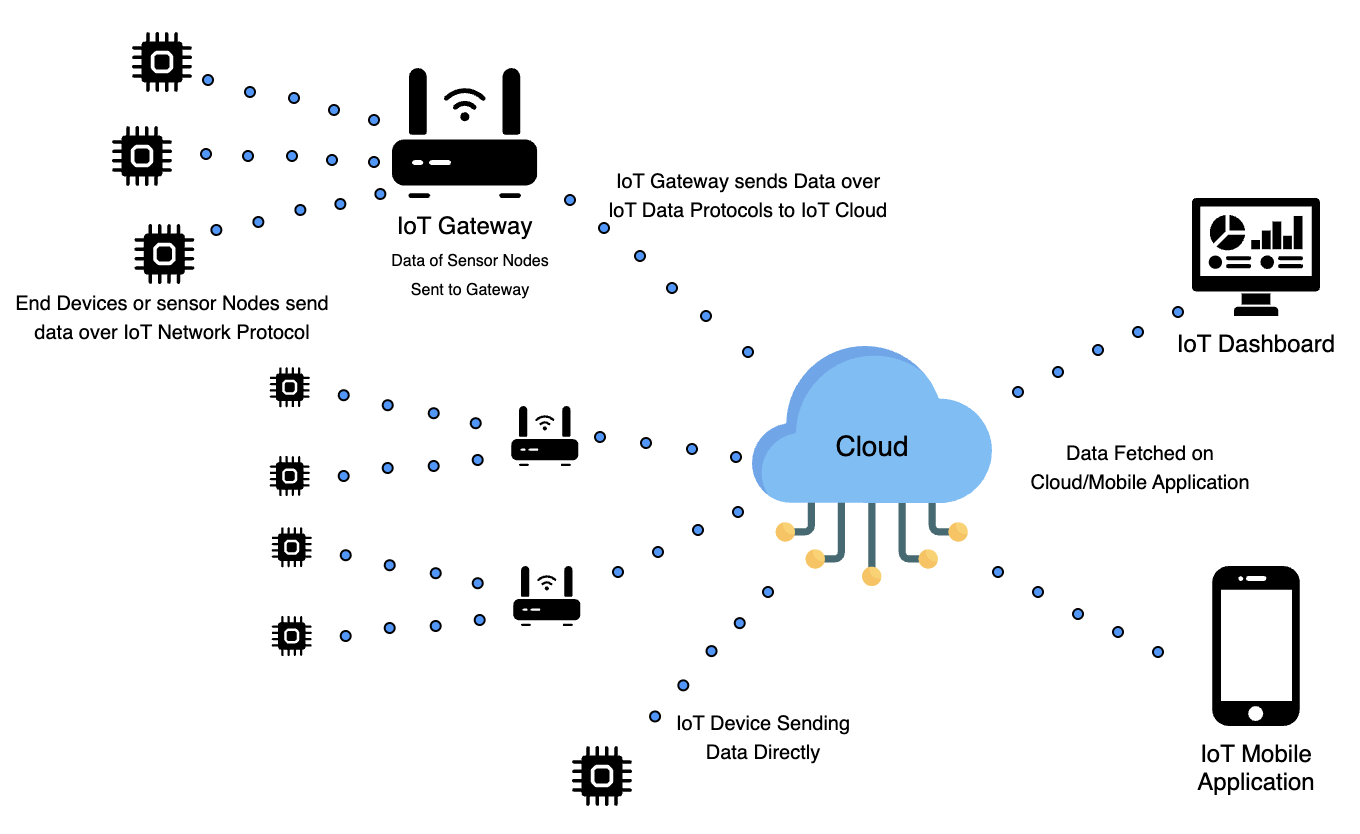}
\caption{
\scriptsize Real-world Cloud-IoT architecture: showing data flows from sensor nodes to the cloud and client interfaces through gateways and IoT protocols.
}
\label{fig:iot_real}
\end{figure}

We construct a cloud-integrated IoT environment using the NS3 network simulator to emulate traffic propagation and DDoS detection. As shown in Figure~\ref{fig:architecture}, a physical cloud server hosts lightweight virtual machines (VMs) representing heterogeneous IoT devices (e.g., sensors, edge nodes, user terminals), which communicate through dynamic NS3-generated topologies emulating CoAP- and MQTT-like traffic patterns. Both normal and DDoS traffic (e.g., SYN flood, UDP reflection) are injected, and real-time traffic logs containing timestamps, IP addresses, and packet sizes are recorded for subsequent detection experiments. Unlike traditional cloud simulators such as CloudSim and OpenStack (shown in Figure~\ref{fig:iot_real}), our framework performs lightweight, traffic-level simulation through the NS3 protocol stack, preserving only key network attributes (IP addresses, ports, traffic rates) while omitting OS and hardware details. This event-driven design supports scalable, reproducible simulations for large-scale deployments involving thousands of devices, with precise control over traffic and attack scenarios.

\begin{figure*}[!ht]
\centering
\includegraphics[width=\textwidth]{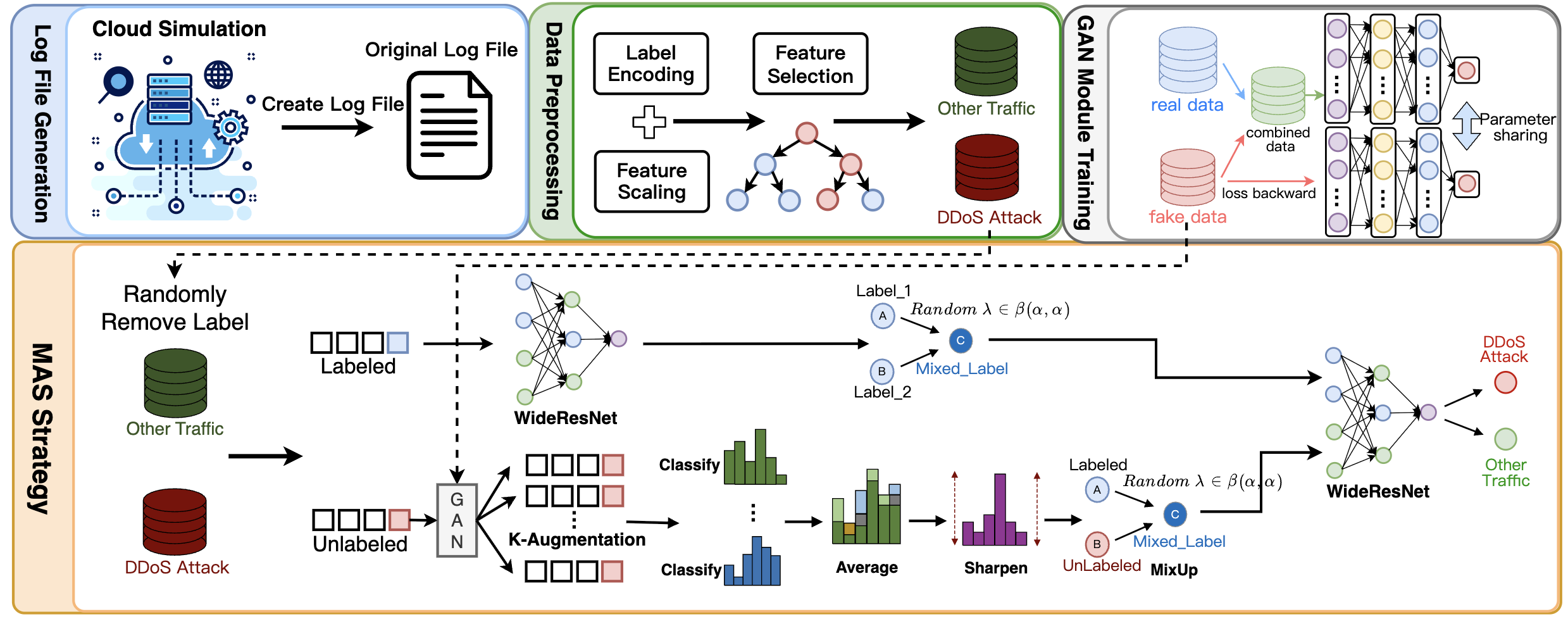}
\caption{
\scriptsize An overview of the MixGAN framework. The process consists of four major stages: 
  (1)~\textbf{Log File Generation:} Synthetic IoT-cloud traffic is generated via NS3 simulation and stored as structured logs; 
  (2)~\textbf{Data Preprocessing:} Raw logs are encoded, scaled, and labeled for downstream classification; 
  (3)~\textbf{CTGAN Training:} A CTGAN module is trained to model minority-class distributions and produce realistic synthetic attack samples; 
  (4)~\textbf{MAS Strategy:} Unlabeled data is augmented $K$ times using CTGAN, pseudo-labeled via WideResNet, then averaged and sharpened. Refined labels are fed into a MixUp module for semi-supervised optimization. The final model jointly learns from labeled and pseudo-labeled traffic.
}
\label{fig:mixgan_framework}
\end{figure*}

\section{Method}

To address the challenges of class imbalance, label scarcity, and complex traffic patterns in IoT-cloud environments, we propose \textbf{MixGAN}, a hybrid detection framework that integrates an enhanced WideResNet backbone, a CTGAN-based augmentation module, and a semi-supervised learning strategy named MixUp-Average-Sharpen (MAS), as illustrated in Figure~\ref{fig:mixgan_framework}.

\subsection{WideResNet for Sequence-Based Feature Learning}

Our backbone adopts a 1-D Wide Residual Network (WideResNet) to capture hierarchical patterns from sequential traffic data. Compared to conventional deep CNNs or RNNs, WideResNet increases the channel width rather than depth, enhancing expressive power while reducing the risk of gradient vanishing and overfitting.

The input $x \in \mathbb{R}^{L}$ is processed by an initial 1-D convolutional layer, followed by three residual blocks with increasing widths (16 $\rightarrow$ 32 $\rightarrow$ 64). Each residual block is defined as:
\begin{equation}
\mathbf{z}^{(i+1)} = \mathbf{z}^{(i)} + \mathcal{F}^{(i)}(\mathbf{z}^{(i)}; \theta_i),
\end{equation}
where
\begin{equation}
\mathcal{F}^{(i)}(z) = \text{Conv}_{1 \times 3}^{(2)} \left( \rho \left( \text{BN}^{(2)} \left( \text{Conv}_{1 \times 3}^{(1)} \left( \rho \left( \text{BN}^{(1)}(z) \right) \right) \right) \right) \right).
\end{equation}

The final output is obtained via global average pooling (GAP) and a fully connected layer:
\begin{equation}
    f_\theta(x) = \sigma(W \cdot \text{GAP}(z^{(B)}) + b).
\end{equation}

\begin{figure}[!ht]
  \centering
  \includegraphics[width=2.8in]{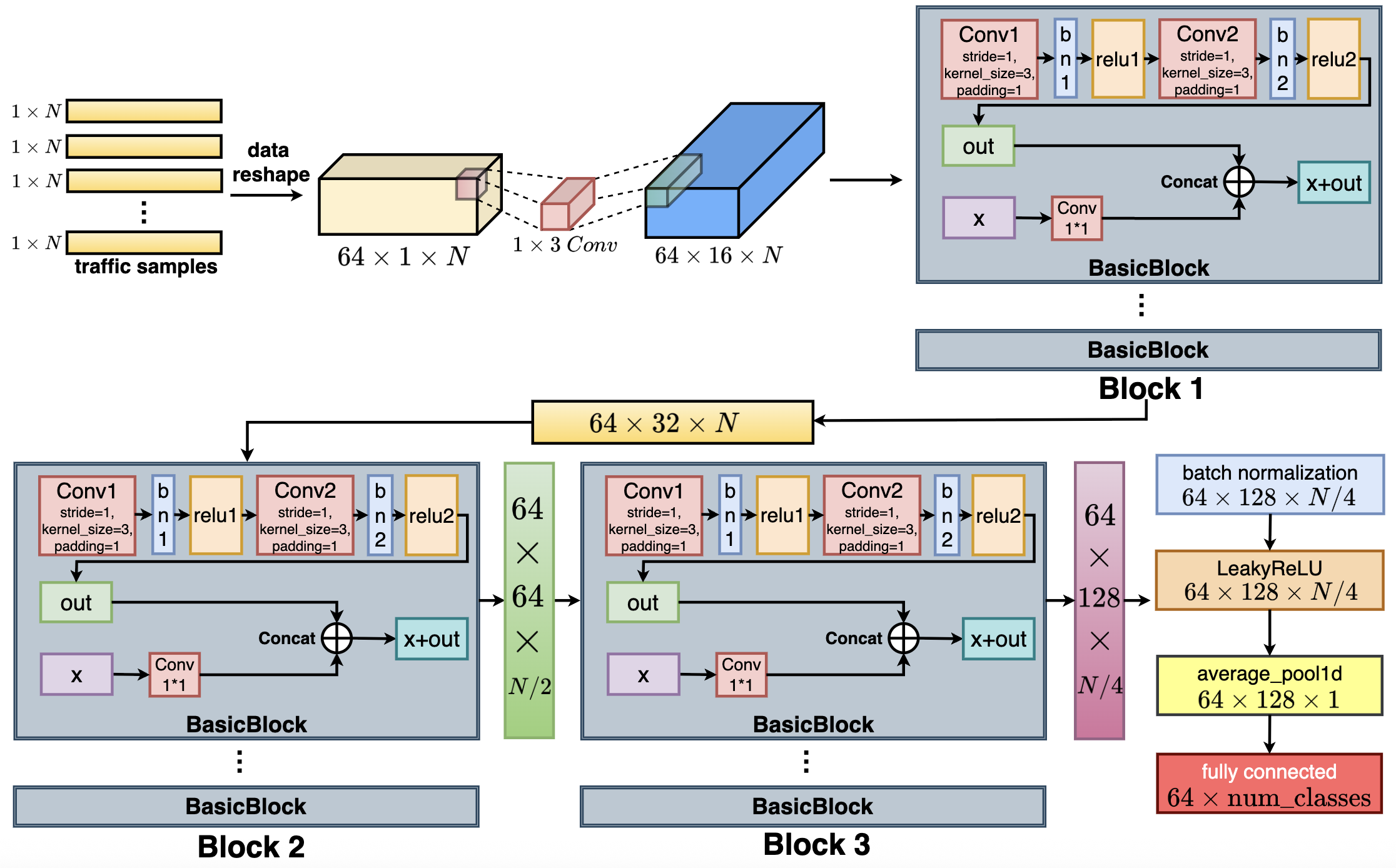}
  \caption{\small Overall architecture of the 1-D WideResNet used in our framework.}
  \label{fig:wideresnet}
\end{figure}

As illustrated in Figure~\ref{fig:wideresnet}, the input traffic sequence is reshaped into a 1-D tensor and passed through three stacked residual blocks. Each block maintains residual connections and expands the channel width while reducing the temporal dimension. The model concludes with average pooling and a fully connected output layer for binary classification.

\subsection{Conditional Data Augmentation with CTGAN}

To alleviate severe class imbalance, we employ a Conditional Tabular GAN (CTGAN) to synthesize high-fidelity attack samples. Let $z \sim \mathcal{N}(0, I)$ be a latent vector and $c \in \{0,1\}$ a class condition. The generator $G(z, c)$ synthesizes a sample, and the discriminator $D(x, c)$ evaluates its realism under the given condition. 

CTGAN's objective can be decomposed into two components: an adversarial loss that estimates the Wasserstein-1 distance between the conditional real and generated distributions, and a gradient penalty to enforce Lipschitz continuity on the discriminator:

\begin{align}
\mathcal{L}_{\text{adv}}(G, D) &= 
\mathbb{E}_{x \sim \mathbb{P}_{\text{real}}} [D(x, c)] 
- \mathbb{E}_{z \sim \mathcal{N}(0, I)} [D(G(z, c), c)] \\[4pt]
\mathcal{L}_{\text{gp}}(D) &= 
\lambda_{\text{gp}} \cdot \mathbb{E}_{\hat{x}} \left[ 
\left( \| \nabla_{\hat{x}} D(\hat{x}, c) \|_2 - 1 \right)^2 
\right],
\end{align}

The full minimax objective is then:

\begin{equation}
\min_G \max_D \; 
\mathcal{L}_{\text{CTGAN}} = 
\mathcal{L}_{\text{adv}} + \mathcal{L}_{\text{gp}}.
\end{equation}

CTGAN enables conditional augmentation of underrepresented classes while modeling both discrete and continuous dependencies in tabular traffic features. The use of conditional inputs ensures that generated samples conform to specific class distributions, thereby improving the diversity and quality of minority-class augmentation.

\subsection{Semi-Supervised Optimization via MAS Strategy}

\begin{table}[htbp]
\centering
\caption{Notations and Definitions in MAS Optimization}
\label{tab:notation}
\renewcommand{\arraystretch}{1.1}
\scriptsize
\begin{tabular}{ll}
\toprule
\textbf{Symbol} & \textbf{Definition} \\
\midrule
$x \in \mathbb{R}^d$, $y \in \{0,1\}$ & Traffic sample and binary label \\
$p_r(x,y) = p_r(y)p_r(x|y)$ & True joint distribution over data \\
$G_\phi(z, c)$ & CTGAN generator, $z \sim p_z$, $c \sim p_r(y)$ \\
$p_g(x,y) = p_r(y)\,[G_\phi(z,c)\,|\,c=y]$ & Distribution of generated samples \\
$\delta = W_1(p_r, p_g)$ & First-order Wasserstein distance \\
$\rho \in (0,1)$ & Ratio of synthetic samples \\
$f_\theta: \mathbb{R}^d \rightarrow \mathbb{R}^2$ & WideResNet classifier outputting logits \\
$\mathrm{softmax}(f_\theta(x))$ & Predicted class probabilities, $p_\theta(y|x)$ \\
$\ell(f_\theta(x), y)$ & Cross-entropy loss; $\ell$ is $L$-Lipschitz \\
$\alpha$ & Beta parameter for MixUp \\
$T \in (0,1]$ & Temperature for label sharpening \\
$\lambda_u$ & Weight for unsupervised consistency loss \\
\bottomrule
\end{tabular}
\end{table}

Inspired by MixMatch, we design a MixUp-Average-Sharpen (MAS) strategy to utilize unlabeled traffic samples, the notations used throughout this section are summarized in Table~\ref{tab:notation}. For each unlabeled sample $u$, $K$ augmented versions are generated using CTGAN and passed through the model to obtain predictions $\hat{p}_1, \dots, \hat{p}_K$. These are averaged and sharpened using temperature $T$ ($T < 1$):
\begin{equation}
    \bar{p}(y|u) = \text{Sharpen}\left( \frac{1}{K} \sum_{k=1}^K \hat{p}_k \right),\ \text{Sharpen}(p)_j = \frac{p_j^{1/T}}{\sum_i p_i^{1/T}}.
\end{equation}

MixUp is applied to both labeled and pseudo-labeled data:
\begin{equation}
    \tilde{x} = \lambda x_1 + (1 - \lambda)x_2, \ 
    \tilde{y} = \lambda y_1 + (1 - \lambda)y_2, \ 
    \lambda \sim \text{Beta}(\alpha, \alpha).
\end{equation}

The final loss is a weighted sum of supervised and unsupervised components:
\begin{equation}
    \mathcal{L} = \mathcal{L}_\text{sup} + \lambda_u \cdot \mathcal{L}_\text{unsup},
\end{equation}

\begin{equation}
    \mathcal{L}_\text{sup} = \sum_{(x, y) \in X_l} \text{CE}(f(x), y),\ 
    \mathcal{L}_\text{unsup} = \sum_{(u, \hat{y}) \in X_u} \|f(u) - \hat{y}\|_2^2.
\end{equation}

This strategy encourages consistency, low entropy, and better generalization in low-label settings typical of real-world DDoS scenarios.

\subsection{Theoretical Analysis of MixGAN}

To theoretically justify the effectiveness of our MixGAN with MAS strategy, we follow a rigorous learning-theoretic analysis framework under mild assumptions, aligning with prior literature on semi-supervised regularization.

\paragraph{Theoretical Justification.}
To explain why MixGAN can generalize well under semi-supervised settings, we analyze its expected risk behavior. Denote real data as $(x,y)\sim p_r(x,y)$, and generated data from a converged CTGAN as $(x,y)\sim p_g(x,y) = p_r(y)[G_\phi(z,c)\mid c=y]$, where $z \sim p_z$ and $c \sim p_r(y)$. Let $\hat{p} = (1-\rho)p_r + \rho p_g$ be the mixed distribution with synthetic injection ratio $\rho\in(0,1)$, and $p_{\text{mix}}$ the MixUp-induced distribution.

Assuming: (A) the classifier $f_\theta$ is $L$-Lipschitz w.r.t. input; (B) CTGAN approximates $p_r$ well with $W_1(p_r, p_g) = \delta \ll 1$; and (C) pseudo labels after temperature sharpening are accurate with probability at least $1-\epsilon$, we aim to upper bound the expected risk on $p_r$.

For the supervised loss $\mathcal{L}_\ell(\theta) = \mathbb{E}_{(x', y') \sim p_{\text{mix}}} \ell(f_\theta(x'), y')$, using the Kantorovich–Rubinstein dual and MixUp smoothing bounds, we obtain:
\begin{align}
    \left| \mathbb{E}_{p_r} \ell - \mathbb{E}_{p_{\text{mix}}} \ell \right|
    &\leq L \cdot W_1(p_{\text{mix}}, p_r) \notag\\
    &\leq L \left( \rho \delta + \tfrac{1}{2} \mathbb{E}_{x_1,x_2} \|x_1 - x_2\|_2 \cdot \mathbb{E}_\lambda |\lambda - 0.5| \right).
    \label{eq:wasserstein_bound}
\end{align}

The Wasserstein term in Eq.~\eqref{eq:wasserstein_bound} regularizes guessed labels by aligning them with true class proportions, reducing overconfidence and mitigating class imbalance. 
This shows that the deviation between training and target risks is tightly controlled by generation quality $\delta$, sample diversity, and MixUp concentration (parameter $\alpha$). For the unsupervised consistency loss $\mathcal{L}_u(\theta)$, we bound the $L_2$ distance between prediction and pseudo-labels via $\epsilon$, leading to:
\begin{equation}
    \mathcal{L}_u(\theta) \leq 4\epsilon.
\end{equation}

Hence, the total expected risk on $p_r$ satisfies:
\begin{equation}
    \mathbb{E}_{p_r} \ell(f_{\hat{\theta}}(x), y) \leq \mathcal{L}(\hat{\theta}) + L\left( \rho \delta + C_\alpha \right) + \lambda_u \epsilon,
\end{equation}

where $\hat{\theta} = \arg\min_\theta \mathcal{L}(\theta)$ and $C_\alpha$ collects the remaining MixUp terms.

This bound reveals that under small $\delta$, $\epsilon$, and $\rho$, the generalization gap is negligible. Moreover, MixUp smooths the empirical distribution, reducing variance and improving robustness. Thus, MixGAN, while injecting synthetic and unlabeled data, does not harm—and may improve—performance under realistic assumptions.

\section{Experiment}
\subsection{Datasets}

To evaluate the performance of the proposed \textbf{MixGAN} framework across diverse network conditions, we utilize three publicly available benchmark datasets: \textbf{NSL-KDD}~\cite{tavallaee2009detailed}, \textbf{BoT-IoT}~\cite{koroniotis2019towards}, and \textbf{CICIoT2023}~\cite{neto2023ciciot2023}.

\textbf{NSL-KDD}~\cite{tavallaee2009detailed} is a refined version of the KDD’99 benchmark, covering DoS, Probe, R2L, and normal traffic, with improved balance and reduced redundancy.  

\textbf{BoT-IoT}~\cite{koroniotis2019towards} emulates real-world IoT networks under diverse attacks (e.g., DDoS, DoS, Reconnaissance, Information Theft) using realistic device and protocol simulations.

\textbf{CICIoT2023}~\cite{neto2023ciciot2023} provides recent IoT traffic traces with multi-vector DDoS, backdoor, and injection attacks, offering fine-grained labels and diverse protocols for evaluating detection in complex IoT-cloud environments.

\subsection{Experimental Setup}

To emulate realistic deployment conditions, we use NS3 to construct cloud-integrated IoT environments corresponding to NSL-KDD, BoT-IoT, and CICIoT2023 datasets. Each environment reproduces typical IoT traffic patterns and attack scenarios from its source dataset, enabling consistent evaluation of MixGAN under diverse network conditions and scales.

For \textbf{NSL-KDD}, 100{,}000 records are randomly sampled. The training set exhibits a 47:1 class imbalance, consisting of 94{,}000 non-DDoS and 2{,}000 DDoS samples, while the test set consists of 2{,}000 non-DDoS and 2{,}000 DDoS records. We simulate 100 IoT nodes with unique IP addresses to replicate a lightweight classical network.

For \textbf{BoT-IoT}, 200{,}000 samples are extracted. The training set includes 170{,}000 non-DDoS and 10{,}000 DDoS samples, corresponding to a 17:1 imbalance ratio. While the test set contains an equal split of 10{,}000 DDoS and 10{,}000 non-DDoS flows. 1{,}000 virtual IoT devices are instantiated to model medium-scale smart environments using protocols such as MQTT and CoAP.

For \textbf{CICIoT2023}, a total of 300{,}000 records are used. The training set comprises 260{,}000 non-DDoS and 20{,}000 DDoS flows, yielding a 13:1 imbalance. The test set again contains 10{,}000 records for each class. A large-scale scenario with 10{,}000 virtual IoT nodes simulates dynamic and heterogeneous communication behavior.

All network interactions are simulated via NS3’s discrete-event engine. Real-time traffic logs capture flow-level features such as timestamp, IP/port, protocol, and packet size, forming the input space for MixGAN training and evaluation.

\subsection{Implementation Details}

To ensure consistency and reproducibility, all experiments are implemented using PyTorch and conducted on a workstation equipped with an NVIDIA RTX 4090 GPU and 24 GB RAM. The overall training pipeline consists of data preprocessing, augmentation, semi-supervised optimization, and evaluation.

\textbf{Data preprocessing and feature engineering.}  
All datasets follow a unified processing pipeline. First, constant-valued features are removed, and missing values are imputed using mean substitution. Categorical attributes (if any) are encoded via one-hot encoding. Next, features are scaled via z-score normalization. We then apply Random Forest-based feature selection to retain the top 15 most important features, ranked using a 70/30 train-validation split. While the selected features remain fixed across datasets, the main experiments adopt varied training ratios for performance evaluation.

\textbf{Backbone architecture.}  
We adopt a 1-D WideResNet-28-2 as the backbone, modified for traffic sequence modeling. The network includes an initial 1-D convolution, followed by three residual blocks with increasing channel widths and downsampling. Each block applies BatchNorm, LeakyReLU, and $1{\times}3$ convolutions. A global average pooling and a fully connected layer are used for classification.

\textbf{CTGAN architecture.} 
Our CTGAN follows the design of~\cite{xu2019modeling}: the generator takes a 128-dimensional noise vector and class label to produce synthetic features, while the discriminator predicts real or fake conditioned on the label. Both networks are implemented as three-layer MLPs with LeakyReLU activations.

\textbf{CTGAN augmentation strategy.} 
We adopt a CTGAN to generate class-conditional synthetic samples for DDoS data. The model is trained offline using a Wasserstein loss with gradient penalty ($\lambda_{\text{gp}} = 10$) for 500 epochs. Across all datasets, CTGAN training remained stable, with average MMD $\approx 0.045$ and W1 $\approx 0.12$ over 500 epochs, indicating well-aligned synthetic and real distributions without mode collapse. We employ CTGAN in two complementary modes, described below.

\begin{table*}[htbp]
\centering
\caption{Complete Performance Metrics Comparison on NSL-KDD and BoT-IoT Datasets}
\label{tab:nsl_bot_comparison}
\scriptsize
\begin{adjustbox}{max width=\textwidth}
\begin{tabular}{ll|cc|cc|cc|cc|cc|cc}
\toprule
\multirow{2}{*}{Metric} & \multirow{2}{*}{Train Ratio} 
& \multicolumn{2}{c|}{LUCID} 
& \multicolumn{2}{c|}{FT-DBN} 
& \multicolumn{2}{c|}{HLBO+DSA} 
& \multicolumn{2}{c|}{FACVO-DNFN} 
& \multicolumn{2}{c|}{GHLBO+DSA} 
& \multicolumn{2}{c}{MixGAN (Ours)} \\
\cmidrule(lr){3-4} \cmidrule(lr){5-6} \cmidrule(lr){7-8} \cmidrule(lr){9-10} \cmidrule(lr){11-12} \cmidrule(lr){13-14}
& & NSL & BoT & NSL & BoT & NSL & BoT & NSL & BoT & NSL & BoT & NSL & BoT \\
\midrule
\textbf{Accuracy} 
& 60\% & 0.806 & 0.797 & 0.884 & 0.863 & 0.857 & 0.885 & 0.904 & 0.887 & 0.899 & 0.903 & \textbf{0.923} & \textbf{0.916} \\
& 70\% & 0.813 & 0.803 & 0.890 & 0.867 & 0.843 & 0.891 & 0.911 & 0.891 & 0.902 & 0.908 & \textbf{0.931} & \textbf{0.929} \\
& 80\% & 0.835 & 0.835 & 0.902 & 0.880 & 0.891 & 0.895 & 0.921 & 0.902 & 0.908 & 0.912 & \textbf{0.943} & \textbf{0.951} \\
& 90\% & 0.847 & 0.841 & 0.909 & 0.890 & 0.897 & 0.899 & 0.930 & 0.914 & 0.914 & 0.917 & \textbf{0.957} & \textbf{0.965} \\
\midrule
\textbf{Precision} 
& 60\% & 0.781 & 0.820 & 0.831 & 0.812 & 0.877 & 0.870 & 0.850 & 0.834 & 0.862 & 0.878 & \textbf{0.914} & \textbf{0.916} \\
& 70\% & 0.801 & 0.826 & 0.837 & 0.815 & 0.865 & 0.876 & 0.856 & 0.837 & 0.871 & 0.885 & \textbf{0.920} & \textbf{0.933} \\
& 80\% & 0.815 & 0.834 & 0.848 & 0.827 & 0.882 & 0.883 & 0.866 & 0.848 & 0.883 & 0.890 & \textbf{0.927} & \textbf{0.953} \\
& 90\% & 0.826 & 0.837 & 0.855 & 0.837 & 0.879 & 0.887 & 0.875 & 0.859 & 0.902 & 0.908 & \textbf{0.944} & \textbf{0.965} \\
\midrule
\textbf{Recall} 
& 60\% & 0.819 & 0.826 & 0.826 & 0.804 & 0.856 & 0.879 & 0.849 & 0.828 & 0.895 & 0.884 & \textbf{0.914} & \textbf{0.916} \\
& 70\% & 0.826 & 0.827 & 0.840 & 0.823 & 0.862 & 0.883 & 0.865 & 0.865 & 0.895 & 0.891 & \textbf{0.926} & \textbf{0.929} \\
& 80\% & 0.832 & 0.832 & 0.863 & 0.846 & 0.865 & 0.889 & 0.883 & 0.879 & 0.907 & 0.896 & \textbf{0.953} & \textbf{0.951} \\
& 90\% & 0.842 & 0.838 & 0.874 & 0.858 & 0.868 & 0.894 & 0.909 & 0.889 & 0.909 & 0.908 & \textbf{0.964} & \textbf{0.965} \\
\midrule
\textbf{F1-Score} 
& 60\% & 0.800 & 0.823 & 0.829 & 0.808 & 0.861 & 0.874 & 0.849 & 0.831 & 0.878 & 0.881 & \textbf{0.914} & \textbf{0.916} \\
& 70\% & 0.813 & 0.827 & 0.839 & 0.819 & 0.860 & 0.879 & 0.861 & 0.851 & 0.883 & 0.888 & \textbf{0.923} & \textbf{0.929} \\
& 80\% & 0.823 & 0.833 & 0.855 & 0.837 & 0.873 & 0.886 & 0.874 & 0.863 & 0.895 & 0.893 & \textbf{0.938} & \textbf{0.950} \\
& 90\% & 0.834 & 0.838 & 0.864 & 0.847 & 0.874 & 0.890 & 0.891 & 0.874 & 0.902 & 0.908 & \textbf{0.952} & \textbf{0.965} \\
\midrule
\textbf{TPR} 
& 60\% & 0.819 & 0.826 & 0.826 & 0.804 & 0.856 & 0.879 & 0.849 & 0.828 & 0.895 & 0.865 & \textbf{0.914} & \textbf{0.916} \\
& 70\% & 0.826 & 0.827 & 0.840 & 0.823 & 0.862 & 0.883 & 0.865 & 0.865 & 0.895 & 0.872 & \textbf{0.926} & \textbf{0.929} \\
& 80\% & 0.832 & 0.832 & 0.863 & 0.846 & 0.865 & 0.889 & 0.883 & 0.879 & 0.907 & 0.878 & \textbf{0.953} & \textbf{0.951} \\
& 90\% & 0.842 & 0.838 & 0.874 & 0.858 & 0.868 & 0.894 & 0.909 & 0.889 & 0.909 & 0.883 & \textbf{0.964} & \textbf{0.965} \\
\midrule
\textbf{TNR} 
& 60\% & 0.771 & 0.819 & 0.857 & 0.832 & 0.879 & 0.884 & 0.893 & 0.854 & 0.857 & 0.872 & \textbf{0.914} & \textbf{0.916} \\
& 70\% & 0.795 & 0.826 & 0.868 & 0.845 & 0.872 & 0.891 & 0.897 & 0.869 & 0.867 & 0.878 & \textbf{0.920} & \textbf{0.933} \\
& 80\% & 0.811 & 0.834 & 0.878 & 0.854 & 0.884 & 0.896 & 0.900 & 0.889 & 0.880 & 0.884 & \textbf{0.927} & \textbf{0.953} \\
& 90\% & 0.823 & 0.837 & 0.889 & 0.875 & 0.881 & 0.908 & 0.929 & 0.900 & 0.901 & 0.909 & \textbf{0.944} & \textbf{0.954} \\
\bottomrule
\end{tabular}
\end{adjustbox}
\end{table*}

\textit{(i) Offline augmentation for ablation.} In ablation studies, we generate tens of thousands of synthetic DDoS samples in advance and mix them into the labeled training set to investigate the effectiveness of pure data-level augmentation.

\textit{(ii) Online augmentation during MAS.} In the proposed MAS pipeline, the pretrained CTGAN serves as a conditional generator to augment unlabeled samples with synthetic variants. These variants are used to construct refined pseudo-labels and are subsequently integrated into MixUp training with labeled samples. The detailed procedure is summarized in the MixGAN workflow below.

\textbf{MixGAN workflow.} In summary, the MixGAN pipeline proceeds as follows:  
(1) generate $K=2$ pseudo-labeled variants for each unlabeled sample via the pretrained CTGAN;
(2) apply sharpening ($T=0.5$) to refine the averaged pseudo-labels; 
(3) perform MixUp with labeled data ($\alpha=0.75$) to promote smooth decision boundaries;  
(4) train the WideResNet backbone on the resulting mixed dataset.

\textbf{Training.} All models are trained for 10 epochs using Adam (learning rate 0.001, batch size 64). For MixGAN, 80\% of labels are masked to simulate a semi-supervised setting, and pseudo-labels for unlabeled samples are generated via the CTGAN-based augmentation pipeline within the MAS framework. The final loss is the cross-entropy over mixed labeled and pseudo-labeled samples, with $\lambda_u = 1.0$.

After training, we report the final performance using accuracy, precision, recall, F1-score, and macro-averaged TPR/TNR.

\subsection{Comparison with State-of-the-art Methods}

To ensure a rigorous and dataset-specific comparison, we select baseline methods that have been recently evaluated on the same datasets in peer-reviewed literature. For the NSL-KDD and BoT-IoT datasets, we include five widely-cited methods: LUCID~\cite{doriguzzi2020lucid}, FT-DBN~\cite{velliangiri2020fuzzy}, FACVO~\cite{gsr2023facvo}, HLBO+DSA~\cite{dehghani2022hybrid} and GHLBO+DSA~\cite{balasubramaniam2023optimization}. These baselines span diverse architectural families, including fuzzy-rule-optimized deep belief networks, hybrid swarm intelligence with stacked autoencoders, and lightweight CNN-based models for efficient DDoS detection in IoT environments.

All methods are evaluated under consistent 10-epoch training settings across multiple train/test splits (60\%–90\%). Full results are reported in Table~\ref{tab:nsl_bot_comparison}, covering accuracy, precision, recall, F1-score, true positive rate (TPR), and true negative rate (TNR).

For the CICIoT2023 dataset, we select recent representative methods that have been evaluated on similar large-scale IoT traffic datasets and are widely used in related intrusion detection studies. These include CLGAN~\cite{li2024hda}, Simple RNN, BiLSTM~\cite{syed2023fog} and a stacking ensemble that integrates Linear SVM, Naïve Bayes, Logistic Regression, and ANN~\cite{khanday2023implementation}. All baselines and our MixGAN model are trained and evaluated using 90\% of the dataset for training. The complete performance is shown in Table~\ref{tab:ciciot2023_comparison}.

\begin{table}[htbp]
\centering
\caption{Performance Comparison on CICIoT2023 Dataset with 90\% Training Data}
\label{tab:ciciot2023_comparison}
\setlength{\tabcolsep}{3pt}
\scriptsize
\begin{tabular}{l|ccccc}
\toprule
\textbf{Metric} & \textbf{Simple RNN} & \textbf{BiLSTM} & \textbf{CLGAN} & \textbf{STACKING} & \textbf{MixGAN (Ours)} \\
\midrule
Accuracy     & 0.903 & 0.878 & 0.909 & 0.919 & \textbf{0.921} \\
Precision    & 0.880 & 0.899 & 0.923 & 0.877 & \textbf{0.925} \\
Recall       & 0.932 & 0.850 & 0.909 & \textbf{0.975} & 0.921 \\
F1-Score     & 0.905 & 0.874 & 0.908 & \textbf{0.924} & 0.920 \\
TPR          & 0.932 & 0.849 & 0.909 & \textbf{0.975} & 0.921 \\
TNR          & 0.873 & 0.905 & 0.923 & 0.864 & \textbf{0.925} \\
\bottomrule
\end{tabular}
\end{table}

\begin{figure*}[!ht]
\centering
\includegraphics[width=\textwidth]{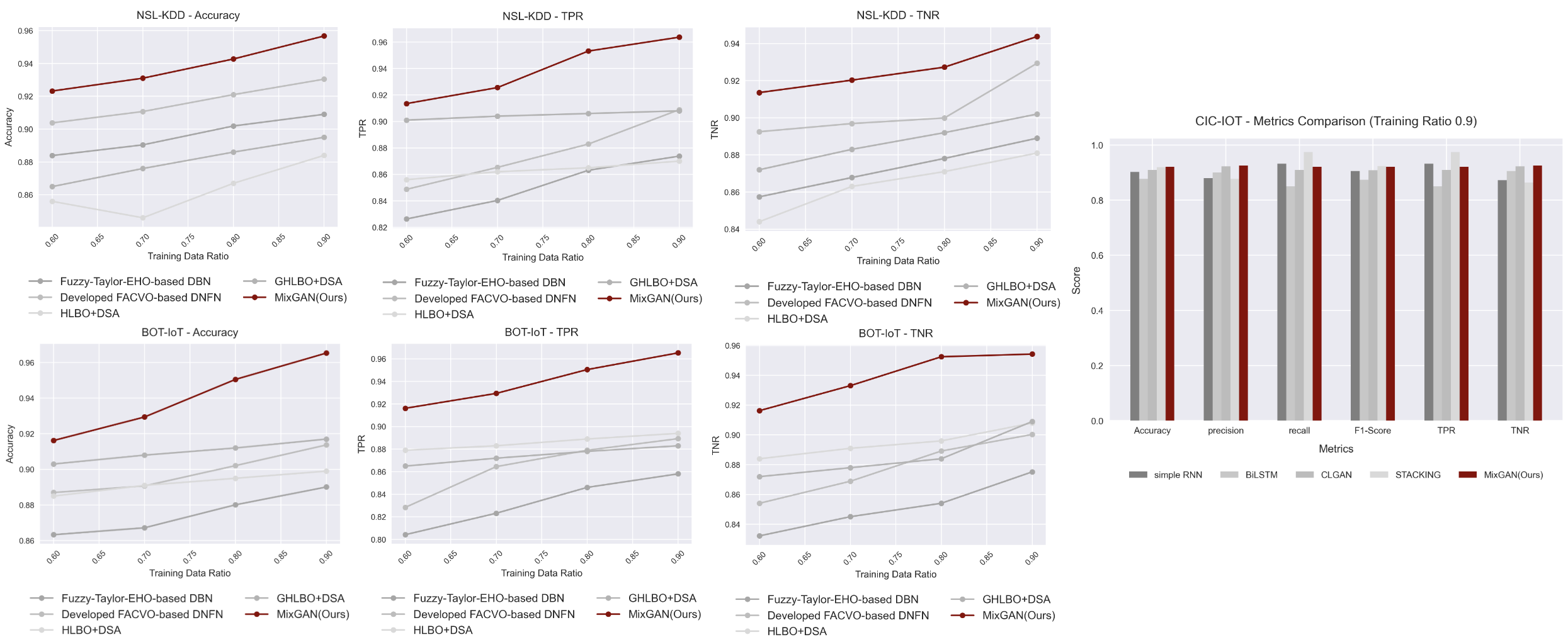}
\centering
\caption{
\scriptsize Performance comparison of MixGAN and baseline methods on NSL-KDD and BoT-IoT datasets under varying training data ratios (60\%–90\%), measured by accuracy, TPR, and TNR. For CICIoT2023, results are evaluated at a fixed training ratio of 90\% and reported using bar plots across six metrics.
}
\label{fig:result}
\end{figure*}

Across all datasets, MixGAN demonstrates superior performance over all baselines (as illustrated in Figure~\ref{fig:result}). On \textbf{NSL-KDD}, it achieves \textbf{95.7}\% accuracy, \textbf{96.4}\% TPR, and \textbf{94.4}\% TNR, outperforming the best baseline by up to 3–5 percentage points. On \textbf{BoT-IoT}, MixGAN reaches \textbf{96.5}\% accuracy, with both TPR and TNR above \textbf{95}\%, highlighting its robustness under noisy traffic conditions. On \textbf{CICIoT2023}, MixGAN attains \textbf{92.1}\% accuracy, 92.5\% precision, and balanced TPR/TNR values of \textbf{92.1}\%/\textbf{92.5}\%, outperforming recent models like CLGAN and stacking model. 


\subsection{Ablation Study}

To better understand the individual contributions of each component in MixGAN, we conduct an ablation study under the 90\% training data setting on NSL-KDD, BoT-IoT, and CICIoT2023 datasets. We compare the following variants:

\begin{itemize}
    \item \textbf{Base:} A vanilla supervised model based on the proposed 1-D WideResNet backbone, trained solely on labeled data without any data augmentation or unlabeled sample utilization.
    \item \textbf{Base + SMOTE:} Applies SMOTE oversampling to the minority class (DDoS attack samples) in the labeled data to address class imbalance.
    \item \textbf{Base + CTGAN:} Incorporates offline-generated samples from CTGAN into the training set to simulate data-level augmentation.
    \item \textbf{Base + MAS:} Employs the MAS strategy (MixUp-Average-Sharpen) on unlabeled data but without using CTGAN-generated samples.
    \item \textbf{MixGAN (Full):} The complete model that combines CTGAN-based conditional generation with the MAS semi-supervised training strategy.
\end{itemize}

\begin{figure}[!ht]
  \centering
  \includegraphics[width=2.8in]{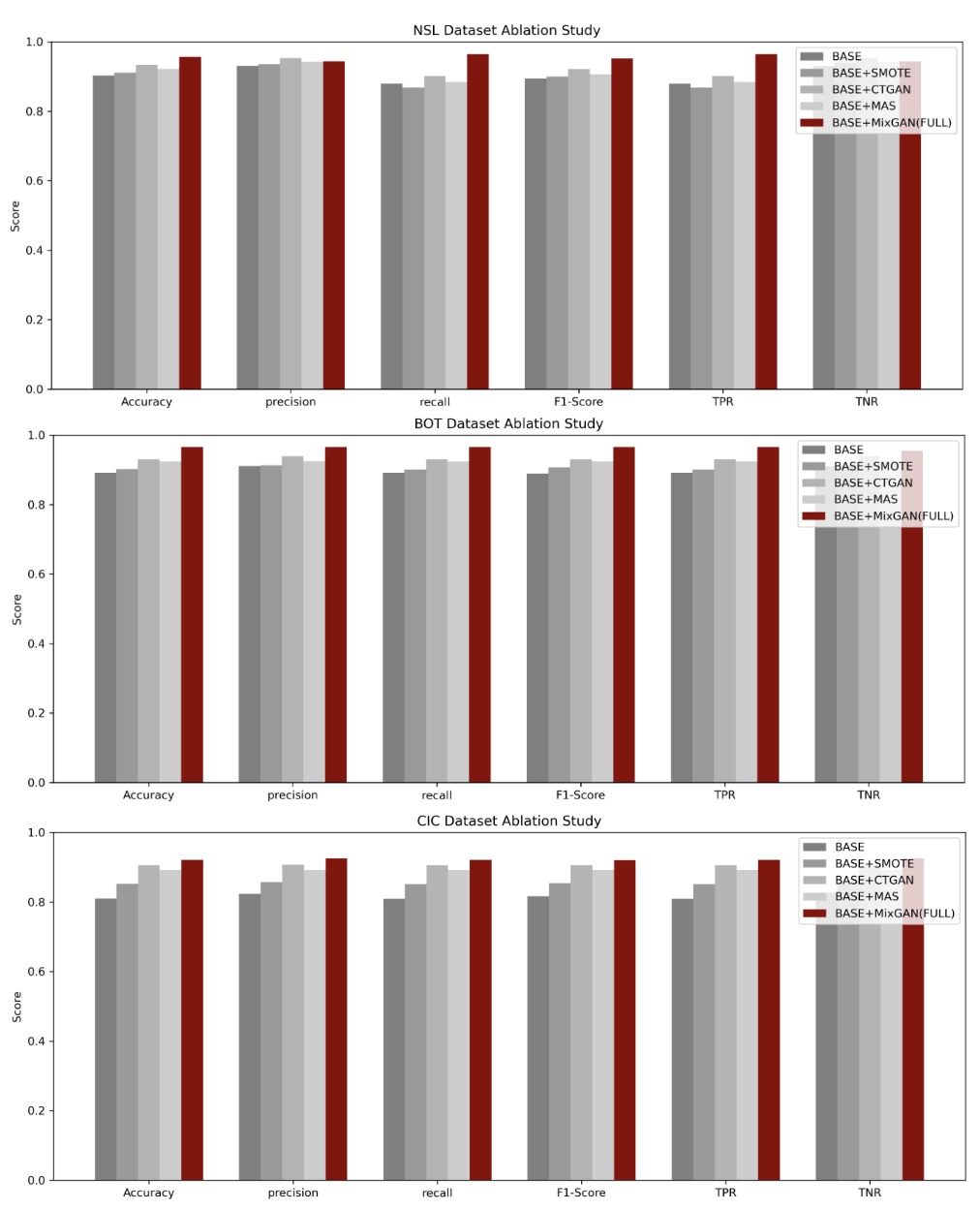}
  \caption{\small Ablation Study on 90\% Training Data.}
  \label{fig:ablation_study}
\end{figure}

All variants are evaluated using accuracy, true positive rate (TPR), and true negative rate (TNR). Table~\ref{tab:ablation_core} summarizes these core metrics, while Figure~\ref{fig:ablation_study} provides a more comprehensive comparison including precision, recall, and F1 score.

\begin{table}[htbp]
\centering
\caption{Ablation Study (Accuracy, TPR, TNR) on 90\% Training Data}
\label{tab:ablation_core}
\scriptsize
\begin{adjustbox}{max width=\linewidth}
\begin{tabular}{l|ccc|ccc|ccc}
\toprule
\multirow{2}{*}{\textbf{Variant}} 
& \multicolumn{3}{c|}{\textbf{NSL-KDD}} 
& \multicolumn{3}{c|}{\textbf{BoT-IoT}} 
& \multicolumn{3}{c}{\textbf{CICIoT2023}} \\
\cmidrule(lr){2-4} \cmidrule(lr){5-7} \cmidrule(lr){8-10}
& Acc & TPR & TNR 
& Acc & TPR & TNR 
& Acc & TPR & TNR \\
\midrule
Base           & 0.903 & 0.879 & 0.931 
               & 0.891 & 0.891 & 0.911 
               & 0.810 & 0.809 & 0.826 \\
Base+SMOTE     & 0.911 & 0.868 & 0.940
               & 0.902 & 0.901 & 0.914 
               & 0.852 & 0.851 & 0.858 \\
Base+CTGAN     & 0.931 & 0.901 & \textbf{0.953} 
               & 0.939 & 0.931 & 0.938 
               & 0.906 & 0.905 & 0.907 \\
Base+MAS       & 0.922 & 0.884 & 0.943
               & 0.925 & 0.924 & 0.925 
               & 0.892 & 0.891 & 0.891 \\
\textbf{MixGAN (Full)} 
               & \textbf{0.957} & \textbf{0.964} & 0.944 
               & \textbf{0.965} & \textbf{0.965} & \textbf{0.954} 
               & \textbf{0.921} & \textbf{0.920} & \textbf{0.925} \\
\bottomrule
\end{tabular}
\end{adjustbox}
\end{table}

\begin{table}[htbp]
\centering
\caption{Effect of Labeling Ratio on BoT-IoT Detection Performance (90\% Training Data)}
\label{tab:label_ratio_bot}
\scriptsize
\begin{tabular}{c|ccc}
\toprule
\textbf{Labeling Ratio} & \textbf{Accuracy} & \textbf{TPR} & \textbf{TNR} \\
\midrule
1\%   & 0.927 & 0.927 & 0.932 \\
5\%   & 0.939 & 0.940 & 0.943 \\
10\%  & 0.955 & 0.955 & \textbf{0.958} \\
20\%  & \textbf{0.965} & \textbf{0.965} & 0.954 \\
\bottomrule
\end{tabular}
\end{table}

\begin{figure}[!ht]
  \centering
  \includegraphics[width=2.8in]{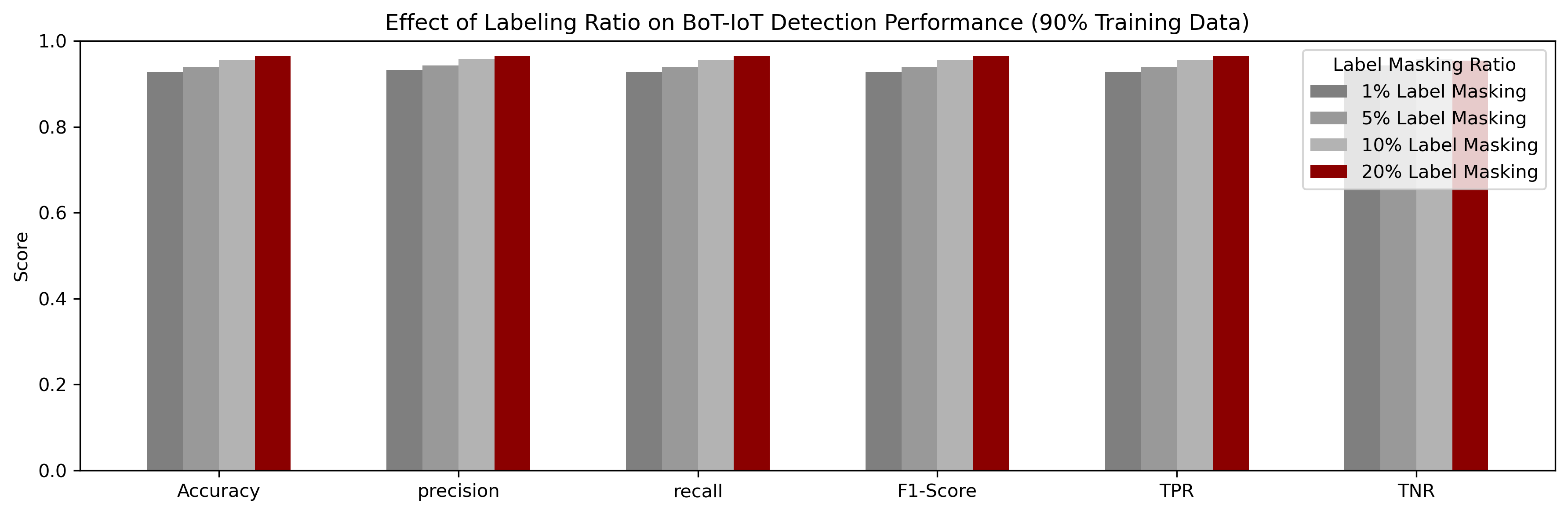}
  \caption{\small Effect of Labeling Ratio on BoT-IoT Detection Performance (90\% Training Data).}
  \label{fig:label_ratio}
\end{figure}

\begin{table}[!ht]
\centering
\caption{Impact of $\alpha$ and $T$ on BoT-IoT Detection Performance (90\% Training Data).}
\scriptsize
\label{tab:mas_hyperparams}
\begin{tabular}{ccc}
\toprule
$\alpha$ & $T$ & Accuracy (\%) \\
\midrule
0.40 & 0.7 & 94.64 \\
0.75 & 0.7 & 94.44 \\
0.90 & 0.7 & 94.66 \\
0.75 & 0.3 & 95.62 \\
\textbf{0.75} & \textbf{0.5} & \textbf{95.80} \\
\bottomrule
\end{tabular}
\end{table}

These results validate the effectiveness of each component in MixGAN. The Base + SMOTE and Base + CTGAN variants improve performance over the baseline by enhancing the sample diversity, while Base + MAS leverages unlabeled data through the pseudo-labeling strategy. However, only the full MixGAN model consistently achieves the highest scores across all datasets and metrics, confirming the benefit of jointly applying CTGAN-based augmentation and MAS-driven semi-supervised optimization.

Besides, to assess robustness under label scarcity, we varied the proportion of labeled data on the BoT-IoT dataset (1\%, 5\%, and 10\%) while keeping the total training size fixed at 90\%. As shown in Table~\ref{tab:label_ratio_bot} and Figure~\ref{fig:label_ratio}, MixGAN sustains high accuracy even with extremely limited labels, and performance degrades only moderately as the labeling ratio decreases. This confirms the MAS strategy’s effectiveness in leveraging unlabeled data for generalization in realistic low-label scenarios.

We further evaluated the sensitivity of MAS to its hyperparameters $\alpha$ (MixUp) and $T$ (sharpening temperature) on BoT-IoT (90/10 split). As shown in Table~\ref{tab:mas_hyperparams}, the MixMatch-style defaults $\alpha=0.75, T=0.5$ deliver the best trade-off, while other settings lead to slight performance drops.

\subsection{Inference Efficiency}
MixGAN is lightweight (3.5MB total) and processes 20k samples in 1.15s ($\approx$0.06 ms/sample) with only 60MB VRAM usage (batch=64) on an RTX 4090. These results indicate strong real-time feasibility for potential edge deployment.

\section{Conclusion}

We propose \textbf{MixGAN}, a generative semi-supervised framework tailored for DDoS detection in IoT-cloud environments. By integrating conditional tabular synthesis with the MAS (MixUp-Average-Sharpen) strategy, MixGAN effectively tackles label scarcity and class imbalance. Experiments on NSL-KDD, BoT-IoT, and CICIoT2023 demonstrate its superiority, achieving up to \textbf{95.7\%} accuracy on NSL-KDD, \textbf{96.5\%} on BoT-IoT, and \textbf{92.1\%} on CICIoT2023, with consistent TPR/TNR gains over state-of-the-art baselines. Comprehensive ablation studies, including component, label-scarcity, and hyperparameter analyses, confirm the robustness and practicality of MixGAN for real-world low-label intrusion detection scenarios.



\begin{ack}
This work is supported by the National Natural Science Foundation of China (No. 62162057, No. 61872254), the Key Lab of Information Network Security of the Ministry of Public Security, China (C20606), the Sichuan Science and Technology Program, China (2021JDRC0004), and the Key Laboratory of Data Protection and Intelligent Management of the Ministry of Education, China (SCUSAKFKT202402Y). We would like to sincerely thank the corresponding author, Jin Yang, for valuable guidance and insightful discussions, which greatly enhanced the quality of this work.
\end{ack}



\newpage
\bibliography{mybibfile}
\end{document}